# THE EXTREME ENERGIES LINES IN THE SOLAR NEUTRINO SPECTRUM

B.I. Goryachev,
Skobeltsyn Institute of Nuclear Physics,
Lomonosov Moscow State University,
119991, Moscow, Russia,
e-mail: bigor@srd.sinp.msu.ru

**Abstract**

Tritium chain of the hydrogen cycle in the Sun including reactions $^3He(e^-,\nu_e)^3H(p,\gamma)^4He$ is considered. At the distance of 1 a.u. the flux of tritium neutrinos is equal to $8.1 \cdot 10^4$ cm$^{-2}$s$^{-1}$. It is an order of magnitude higher than the flux of the (hep)-neutrinos. Radial distribution of $^3H$-neutrinos yield inside the Sun and their energy spectrum which has a form of line at the energy of $(2,5 \div 3,0)$ keV are calculated. The flux of thermal tritium neutrinos is accompanied by a very weak flux of antineutrinos ($\sim 10^3$ cm$^{-2}$year$^{-1}$) with energy lower than 18,6 keV. These antineutrinos are produced during Urca processes $^3He \leftrightarrows ^3H$. The flux of the neutrinos of maximum possible energy (line 19,8 MeV) produced due to the (heep)-reaction (related to the (hep)-process) is estimated.

## 1. Introduction

The Standard Solar Models (SSM) which describe in details the processes of the neutrino production are in a good mutual agreement (for instance [1], [2]). Reactions of hydrogen cycle covering the yield of neutrinos up to two percent accuracy are of particular interest.

It is well known that these reactions are started from the processes

$$p + p \to {}^2H + e^+ + \nu_e, \qquad (1)$$

$$^2H + p \to {}^3He + \gamma \qquad (2)$$

and then are separated into several chains.

All basic (including not more than two particles in the initial state) chains of hydrogen cycle and a number of their characteristics are presented in Table. In the works on SSM four from five chains (№№ 1, 2, 3 and 5) are described.

In Section 2 tritium chain which gives not the least contribution to the neutrino flux is studied.

The rate of tritium production due to solar plasma electron capture under conditions in the center of the Sun is approximately estimated in [3].

The results of the calculation of neutrino flux in $^3H$-chain and the distribution of tritium neutrino yield along the radius of the Sun are presented below.

Energy spectrum of these neutrinos which is directly determined by the temperature in the solar interior is calculated.

The flux of electron antineutrino produced by Urca processes associated with $^3H$-chain is determined.

The flux of the neutrinos from the (heep)-reaction which is a three-body process related to the basic (hep)-reaction is calculated in Section 3. The most energetic solar neutrinos (with energy of 19,8 MeV) are produced in the (heep)-reaction.

SSM's characteristics obtained in [1] are used in the calculations.

## 2. Tritium chain of hydrogen cycle.

In $^3H$-chain the reaction (2) is followed by the capture of the electron from solar plasma

$$^3He + e^- \to {}^3H + \nu_e, \qquad (3)$$

which results in tritium nuclei production.

This chain is completed with fusion reaction

$$^3H + p \to {}^4He + \gamma, \quad E_\gamma = 19{,}8\ MeV. \qquad (4)$$

Unlike all other reactions of the hydrogen cycle (as well as CNO cycle) the process (3) is an endoergic process with $E_{thr}$ threshold which is equal to the maximum energy of the electrons produced by β-decay of tritium, i.e. 18,6 keV.

Consistent with the Fermi "golden rule" and neglecting the input of heavy elements (z>2) into the electron density the following expression for the rate of tritium nuclei production per one $^3He$ nucleus we'll obtain:

$$R(s^{-1}) = (4\pi m_p \hbar^4 c^3)^{-1} \cdot G^2 \cdot \left|\tilde{M}\right|^2 \rho(1+X) \cdot \left\langle (E_e - E_{thr})^2 \cdot F(E_e, z) \right\rangle, \qquad (5)$$

where pointed brackets mean averaging over the thermal energy electron spectrum $E_e$. In (5) $G$, $F(E_e, z=2)$ and $m_p$ specify Fermi constant, Fermi function and proton's mass, respectively. The other designations are standard for SSM. The reduced matrix element $\tilde{M}$ corresponding to β-decay of tritium is taken equal to analogous element for β-decay of neutron because within the limits of the possible accuracy of estimation they do not differ [4].
Let

$$\left\langle (E_e - E_{thr})^2 \cdot F(E_e, z=2) \right\rangle \equiv 2\pi^{-1/2}(kT)^{-3/2} I_1(E_e, E_{thr}, T), \qquad (6)$$

where

$$I_1 = \int_{E_{thr}}^{\infty} e^{-\frac{E_e}{kT}} E_e^{1/2} (E_e - E_{thr})^2 F(E_e, z) dE_e. \qquad (6a)$$

As seen from (6) and (6a) averaging over the energy spectrum of the electrons uses Boltzman-Maxwell distribution which describes the "tail" of the real thermal distribution properly [5]. Turning to the energy $E_\nu$ of the neutrinos produced in (3) we'll get

$$I_1(E_\nu, E_{thr}, T) = e^{-\frac{E_{thr}}{kT}} E_{thr}^{1/2} I_2(E_\nu, E_{thr}, T) \qquad (7)$$

where

$$I_2(E_\nu, E_{thr}, T) = \int_0^\infty e^{-\frac{E_\nu}{kT}} [1 + (E_\nu / E_{thr})]^{1/2} E_\nu^2 F(E_\nu + E_{thr}, z) dE_\nu \qquad (7a)$$

An analytic approximation formula [6] convenient both for nonrelativistic and for relativistic energies $E_e$ was used for the Fermi function $F(E_e, z)$. In the $E_e$ interval of interest function $F(E_e, z)$ varies slightly and at the mean is equal to 1,15. Integrating over $E_\nu$ and taking into consideration (5)-(7a), we'll obtain

$$R(s^{-1}) = 4{,}1 \cdot 10^{-16} \rho(1+X)(kT)^2 e^{-\frac{E_{thr}}{kT}} \left[ 2{,}52 \left(\frac{E_{thr}}{kT}\right)^{1/2} + 7{,}3 \cdot 10^{-2} \left(\frac{m_e c^2}{kT}\right)^{1/2} \right] \qquad (8)$$

In formula (8) all energies and $kT$ values must be expressed in units of keV. Expression (8) is true for the fixed radius $r$ inside the Sun (in SSM it means $T=const$). Averaging over the region of neutrino production in the reaction (3) gives the characteristic lifetime $\tau$ of $^3He$ nucleus in this reaction

$$\tau \equiv R^{-1} \cong 1{,}4 \cdot 10^{11} \text{ years} \qquad (9)$$

It is an order of magnitude lower than the analogous value for the (hep)-reaction [7, 8].

Specific yield of the tritium neutrinos $W(cm^{-3}s^{-1})$ is

$$W(cm^{-3}s^{-1}) = N_A R(s^{-1}) \rho [X(^3He)/3] \qquad (10)$$

Integrating over the Sun's volume we'll obtain the flux $F$ of the tritium neutrinos at a distance of 1 a.u.

$$F_\nu = 8{,}1 \cdot 10^4 \, cm^{-2} s^{-1}, \qquad (11)$$

which is an order of magnitude higher than the corresponding flux for the (hep)-reaction (see Table).

An advantage of the production of tritium nuclei (3) in reference to the (hep)-reaction is that it's not connected with overcoming of the Coulomb barrier. In the reactions of electrons' capture Coulomb interaction only increases the capture cross-section $(F(E_e,z)>1)$. But in the endoergic process (3) another exponential factor of smallness appears instead of Gamow penetration coefficient. It is a threshold factor $exp(-E_{thr}/kT)$. For all possible threshold capture reactions in the interior of the Sun for light and medium nuclei this factor is vanishing and substantially it leads to the zero cross section for these reactions.

For the concerned process (3), where $E_{thr}=18{,}6 \, keV$ the threshold is extremely low. It provides essentially higher flux of the neutrino in tritium chain than in the (hep)-reaction. It's worthy of note that according to SSM the concentration of $^3He$ nuclei (in distinction from the other lightest nuclei with $z>1$) grows up to the radius $r \approx 0{,}27 \, R_\odot$. It also contributes to the growth of the flux of tritium neutrinos.

Radial distribution of the yield of tritium neutrinos calculated with regard to (8) and (10) is presented at Fig.1. It is close to the analogous relative distribution of the yield of neutrinos in $^7Be$-chain known in SSM.

In principle, capture (3) can be followed with β-decay of tritium. Then twin Urca processes $^3He \leftrightarrows \, ^3H$ resulting in the production of the solar antineutrino will take place. Without reaction (4) this flux would be equal to the flux $F_\nu$ in (11). The conditions of β-decays realization in the interior of the degenerate and undegenerate stars were studied by a number of authors (for instance, see [9]). In case of the Sun these conditions substantially do not differ from the Earth conditions. Therefore the energy of the antineutrinos will not exceed 18,6 keV, and lifetime of tritium nucleus relative to β-decay is $\tau_\beta = 17,7$ years. The calculated value of the lifetime $\tau_{tn}$ of $^3He$ nucleus relative to nuclear fusion (4) is $\tau_{tn} \cong 0,24$ s. As a result taking into consideration (11) the flux of the solar antineutrinos at a distance of 1 a.u. is

$$F_{\bar{\nu}} = F_\nu (\tau_{tn}/\tau_\beta) \cong 10^3 \, cm^{-2} \, year^{-1}$$

Energy distribution of tritium neutrinos at fixed temperature $T\,(r=const)$ is given by integrand in (7a). Neutrino energy spectrum after averaging over the Sun's volume is presented at Fig.2. Width of this distribution is comparable with the width of the lines in the spectrum of the solar neutrinos known in SSM. Therefore it's possible to tell that the yield of tritium neutrinos forms the line at the energy of $E_\nu \cong (2,5 \div 3,0) keV$, and the neutrinos themselves are naturally to be named "thermal".

This circumstance fundamentally set $^3H$-neutrino apart from the neutrino produced by the other chains of hydrogen cycle, as well as CNO-cycle, where neutrino energy scale is determined by mass relations.

As seen from (7a) the maximum of the energy spectrum of the neutrinos produced in the region with temperature T corresponds to the energy $E_\nu^*$,

$$E_\nu^* \cong 2kT .\qquad(12)$$

So the position of the line of tritium neutrinos in the averaged energy spectrum according to (12) makes it possible to determine effective temperature of the zone of their intensive production. It's possible to show that the input of the thermal neutrinos' line is about 40% of the yield of the neutrinos with continuous energy spectrum in the region of the line. Therefore if we could at least roughly measure neutrino spectrum in the thermal region then determination of the concerned line's position would give a possibility for the direct thermometry of the solar interior.

As is known for all the chains of hydrogen cycle the result of the chain of step-by-step reactions can be expressed as following:

$$4p + 2e^- \rightarrow {}^4He + \nu_{e1} + \nu_{e2} + Q .\qquad(13)$$

In (13) $\nu_{e1}$ – neutrino produced in (1), and $\nu_{e2}$ – neutrino produced in specific for each chain of reaction (for $(^3He+^3He)$-chain $\nu_{e1}=\nu_{e2}$), $Q = 26,7$ MeV. The mean thermal energy <Qw> released in each chain is presented in Table 1 along with other characteristics of hydrogen cycle chains. It's obtained by subtraction of mean energy values for $\nu_{e1}$ and $\nu_{e2}$ neutrinos from Q. As seen from Table this value is maximum for tritium chain.

### 3. The (heep)-reaction as a source of neutrinos with maximum energy.

It is believed that neutrinos with maximum energy are produced in the (hep)-reaction

$$^3He + p \rightarrow {}^4Li \rightarrow {}^4He + e^+ + \nu_e, \qquad (14)$$

and the value of energy is equal to maximum energy of β-decay $^4Li$, i.e. 18,77 MeV. But in fact neutrino produced in the reaction

$$^3He + e^- + p \rightarrow {}^4He + \nu_e, \qquad (15)$$

must provide the maximum energy. This reaction is related to (14). A line with energy 19,8 MeV is produced in this reaction. Taking into consideration that reaction related to pp-process (1) is usually denoted as (pep), it is natural to denote (15) as (heep)-reaction.

Apparently the (heep)-reaction is the single reaction whose input into solar neutrino yield can be measured separately. Therefore it is interesting to estimate the (heep)-neutrino flux. For the reaction of electron capture similar (15) with a three-body initial state the following relation is true:

$$W(heep) = K(E_\nu^*, T, z, n_e) f^{-1} W(hep), \qquad (16)$$

where *W(heep)* and *W(hep)* – are yield of corresponding reactions (s$^{-1}$cm$^{-3}$), *f* – non-dimensional factor of phase volume for β-decay of $^4Li$, *K* – analogous generalized factor for electron capture in (15). In (16) $E_\nu^*$ marks the energy of the (heep)-neutrino line, *T* and $n_e$ –temperature and density of the electrons in the solar zone with radius *r*, respectively, and *z=3* is total charge of hadrons in the initial state (15). Calculations give the following value for f:

$$f = 2,4 \cdot 10^6 \qquad (17)$$

For the case of arbitrary light nuclei the functions *K($E_\nu$\*, T, z, $n_e$)* are presented in [10]. Following [10] and taking into consideration (16) and (17) we'll obtain the following relation for the fixed radius *r*:

$$W(heep)/W(hep) = 8,13 \cdot 10^{-10} \rho (1+X) T_6^{-1/2} (1+0,02 T_6), \qquad (18)$$

where $T_6$ is temperature in the units of $10^6$ K.

Relation (18) is five orders of magnitude smaller than *W(pep)/W(pp)* obtained in [10]. Basically it's explained by an extremely high factor (17), which in turn is associated with unusually high value of maximum energy of β-decay of $^4Li$. Taking into consideration (18) and known from SSM values of neutrino flux and radial distribution of neutrino yield in the (hep)-reaction we'll obtain the following value for the (heep)-neutrino flux at a distance of 1 a.u.

$$F_\nu(heep) = 2,5 \cdot 10^{-4} cm^{-2} s^{-1} \qquad (19)$$

Note, that it is possible to calculate the relations similar to *W(pep)/W(pp)* and *W(heep)/W(hep)* reasonably accurately, because these calculations are basically associated with accounting of weak interaction processes kinematics and do not require information about the cross-sections of nuclear reactions. Therefore measuring the flux *$F_\nu$(heep)* would provide a principle possibility to clarify the flux *$F_\nu$(hep)*. Evaluation of *$F_\nu$(hep)* by different authors [7,8] varies greatly. Unfortunately the flux (19) is extremely small.

## 4. Conclusion

Accounting of tritium chain provides an opportunity to obtain full description of the basic chains of hydrogen cycle on the Sun within the framework of SSM. The flux of tritium neutrinos at a distance of 1 a.u. is $8,1 \cdot 10^4$ cm$^{-2}$s$^{-1}$, and it is an order of magnitude greater that analogous flux of the (hep)-neutrinos. Radial distribution of the yield of $^3H$-neutrinos in relative units is close to the similar distribution of $^7Be$-neutrinos. Tritium neutrinos form a line of thermal neutrinos with energy of (2,5÷3,0) keV. Since reaction $^3He(e^-,\nu_e)^3H$ is endoergic one, its rate is strongly suppressed by exponential threshold factor $-\exp(-E_{thr}/kT)$.

Due to the possible decay of tritium along with the reactions of tritium chain Urca processes $^3He \leftrightarrows ^3H$ accompanied by production of antineutrinos with energy of $E_{\bar{\nu}} < 18,6\ keV$ go with a very low probability. The flux of antineutrino is about $10^3$ cm$^{-2}$year$^{-1}$.

The maximum possible energy of the solar neutrinos is 19,8 MeV. These neutrinos form an energetic line and are produced as a result of the (heep)-reaction – related to the (hep)-process. The flux of the neutrinos of such energy is $2,5 \cdot 10^{-4}$ cm$^{-2}$s$^{-1}$.

## Acknowledgements

The author is grateful to V.I. Galkin and S.I. Svertilov for a fruitful discussion.

**Table. Characteristics of the basic chains of hydrogen cycle.**

The Table presents the reactions following (1) and (2). The flux of neutrinos $F_\nu$ are given at a distance of 1 a.u. Values $E_\nu$ correspond to the maximum energies of neutrinos in continuous spectra and to the lines in the case of electron capture. In the last column the mean thermal energy <Qw> released in each chain is presented.

The characteristics of $^3H$-chain are obtained in the present paper.

| № | chain | $F_\nu$, cm$^{-2}$s$^{-1}$ | $E_\nu$ | <Qw>, MeV |
|---|---|---|---|---|
| 1 | ($^3$He+$^3$He)–**chain** $^3$He($^3$He,2p)$^4$He | 6,1·10$^{10}$ | <0,42 MeV | 26,2 |
| 2 | $^7$Be–**chain** $^3$He($^4$He,γ)$^7$Be(e$^-$,ν$_e$)$^7$Li(p,γ)2$^4$He | 4,7·10$^9$ | 0,862 MeV (90%) 0,384 MeV (10%) | 25,6 |
| 3 | $^8$B–**chain** $^3$He($^4$He,γ)$^7$Be(p,γ)$^8$B $^8$B→2$^4$He+e$^+$+ν$_e$ | 5,8·10$^6$ | <14,6 MeV | 19,7 |
| 4 | $^3$H–**chain** $^3$He(e$^-$,ν$_e$)$^3$H(p,γ)$^4$He | 8,1·10$^4$ | (2,5÷3,0) keV | 26,4 |
| 5 | (hep)–**chain** $^3$He(p, e$^+$ν$_e$)$^4$He | 8·10$^3$ | <18,8 MeV | 16,8 |

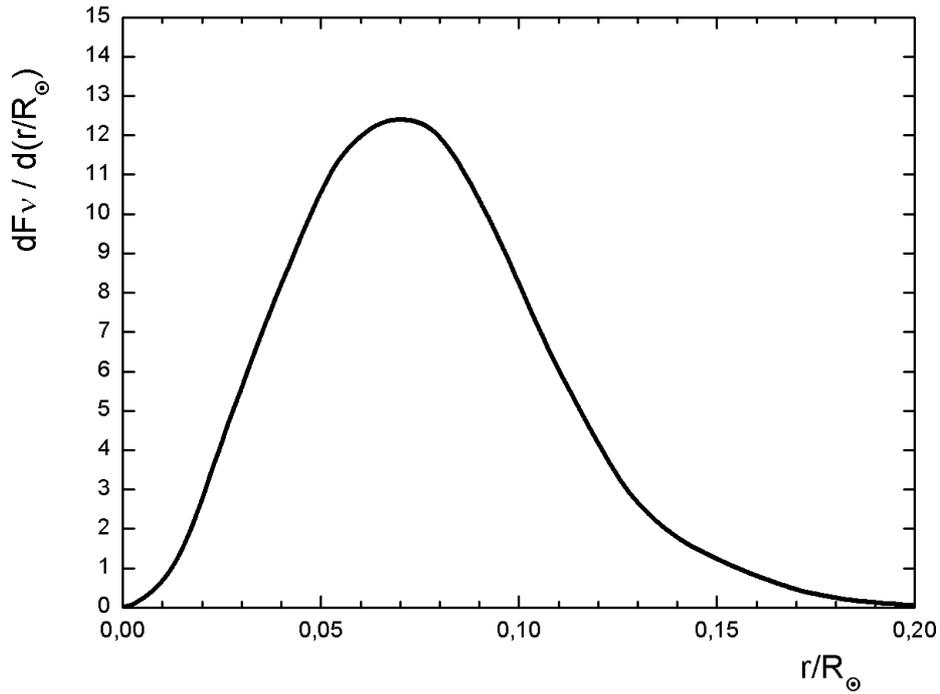

**Fig.1.** Differential radial distribution of the flux (yield) of tritium neutrinos. Area under the curve is normalized.

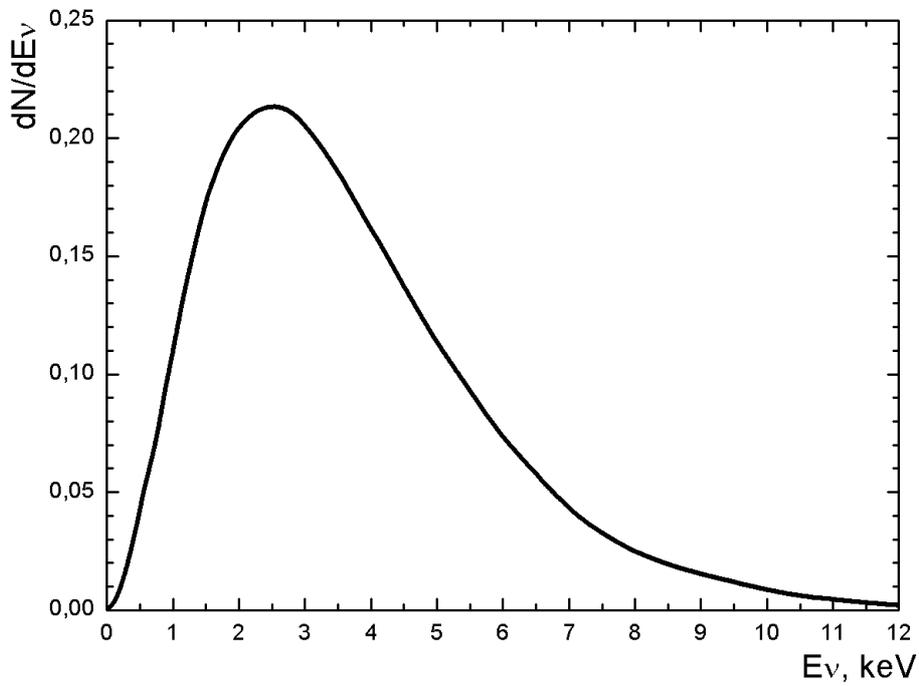

**Fig.2.** Differential energy spectrum of tritium neutrinos. When measuring the energy of neutrino in keV the area under the curve is normalized to unity.